\documentclass[aps,
floatfix,superscriptaddress,a4paper]{revtex4}
\usepackage{graphicx,amsmath,bbm,mathrsfs,amssymb,psfrag}

\evensidemargin \oddsidemargin
\begin{document}

\author{Laure Gouba}

\affiliation{
The Abdus Salam International Centre for
Theoretical Physics (ICTP),
 Strada Costiera 11,
I-34151 Trieste Italy.
Email: lgouba@ictp.it}

\title{Dirac's Method for the Two-Dimensional Damped Harmonic Oscillator in the Extended Phase Space }

\date{\today}

\begin{abstract}
The system of two-dimensional damped harmonic oscillator is revisited in the extended phase space.
It is an old problem already addressed by many authors that we present here in some fresh points 
of view and carry on smoothly a whole discussion. 
We show that the system is singular. The classical Hamiltonian  is proportional 
to the first-class constraint. We pursue with the Dirac's canonical quantization procedure 
by fixing the gauge and provide a reduced phase space description of the system.
As result the quantum system is simply modeled by the original quantum Hamiltonian. 
\end{abstract}

\maketitle

\section{Introduction}\label{sec1}

The Hamiltonians of most real physical systems are explicitly time dependent 
and do not provide directly conserved quantities. Succeeding to isolate an 
invariant helps to know about a fundamental system property. For instance in the 
case of an autonomous Hamiltonian system the Hamiltonian itself represent an invariant.
Many approaches have been developed 
to identify conserved quantities for explicitly time-dependent systems. The first one
being developed by Emmy Noether in the context of the Lagrangian formalism \cite{emmy}.
The invariant for the one-dimensional time-dependent harmonic oscillator has been 
derived by H. R. Lewis \cite{lewis}. It has been demonstrated later that the Lewis procedure 
follows from Noether Theorem \cite{lutzky} and that has been extended by Chattopadhyay 
to derive invariants for certain one-dimensional non-linear systems \cite{padhyay}.
An other approach of finding conserved quantities for explicitly time-dependent has 
been developed by Leach by performing a finite time dependent canonical transformation 
\cite{leach}. A third way of finding exact invariants for time-dependent classical Hamiltonians
has been derived by Lewis and Leach by using direct {\it Ans\"atze} with different powers in 
the canonical momentum \cite{lewlea}.

The invariants for time-dependent Hamiltonian systems are still investigated 
and of interest in the literature. We have been first interested in 
finding the class of invariants for the two-dimensional time-dependent Landau problem 
and harmonic oscillator in a magnetic field \cite{laure} where we considered an isotropic two-dimensional harmonic 
oscillator with arbitrarily time-dependent mass $M(t)$ and frequency $\Omega(t)$ in an arbitrarily time-dependent 
magnetic field $B(t)$. Two commuting invariant observables 
(in the sense of Lewis and Riesenfeld) $L$, $I$  have been derived in terms of some solutions of an auxiliary ordinary 
differential equation and an orthonormal basis of the Hilbert space consisting of joint eigenvectors 
$\varphi_\lambda$ of $L$, $I$. 

Recently we studied a system of two non-interacting damped oscillators 
with equal time-dependent coefficients of friction and equal time-dependent frequencies \cite{latevi}.
The system is described by the Lagrangian function

\begin{equation}\label{lagrr}
L(x_1,x_2,\dot{x}_1,\dot{x}_2,t) = f^{-1}(t)
\left(\frac{m}{2}(\dot{x}_1^2 +\dot{x}_2^2) -\frac{m\omega^2(t)}{2}(x_1^2 +x_2^2)\right),
\end{equation}

where $f$ is an arbitrary function such that $f(t) = e^{-\int_0^t\eta(t')dt'}$, we assume that 
the function $f$ is twice differentiable and the canonical coordinates are $x_1, x_2$. The 
canonical momenta are respectively given by 

\begin{eqnarray}
 p_1 &=& \frac{\partial L}{\partial\dot{x}_1}  = m f^{-1}(t)\dot{x}_1;\\
 p_2 &=& \frac{\partial L}{\partial\dot{x}_2}  = mf^{-1}(t)\dot{x}_2.
\end{eqnarray}

In the canonical formalism the dynamics of the system is govern by the classical Hamiltonian 

\begin{equation}
 H(x_1,x_2,p_1,p_2, t) = p_1\dot{x}_1 + p_2\dot{x}_2 - L(x_1,x_2,\dot{x}_1,\dot{x}_2,t), 
 \end{equation}
 
 that is equivalent to
 
 \begin{equation}\label{hamm}
 H(x_1,x_2,p_1,p_2, t) = \frac{f(t)}{2m}(p_1^2 + p_2^2) + f^{-1}(t)\frac{m\omega^2(t)}{2}(x_1^2 +x_2^2).
\end{equation}

The dynamics of the system are determined by the values of the canonical coordinates 
and momenta at any given time $t$. The coordinates and momenta satisfy a set of Poisson Brackets
relations

\begin{equation}\label{pbracket}
 \{x_1, p_1\}_{PB} =1; \quad \{x_2, p_2\}_{PB} = 1; \quad \{x_1, p_2\}_{PB} = \{x_2, p_1\}_{PB} = 0, 
\end{equation}

and

\begin{equation}
\{f, g\}_{PB} =\frac{\partial f}{\partial x_1}\frac{\partial g}{\partial p_1}  +
\frac{\partial f}{\partial x_2}\frac{\partial g}{\partial p_2 } 
- \frac{\partial f}{\partial p_1}\frac{\partial g}{\partial x_1}  -
\frac{\partial f}{\partial p_2}\frac{\partial g}{\partial x_2 }\;,
\end{equation}

where $f$ and $g$ are any functions of the $x_i$'s and $p_i$'s, $i =1,2$.
At the quantum level the dynamic invariant method formulated 
by  Lewis-Riesenfeld \cite{lewisr} has been used to construct an exact invariant operator.
The exact solutions for the corresponding time-dependent Schr\"odinger equations are provided. 
The solutions have been used to derive the generators of the $su(1,1)$ Lie algebra that enable to 
construct and study the properties of the coherent states {\it $\grave{a}$ la } Barut-Girardello 
and {\it $\grave{a}$ la } Perelomov.

In this paper we revisit the model in \cite{latevi} in the extended phase space. The idea 
is not new in the literature. For instance we refer to the following works that we met in the 
literature : the one by Struckmeier on 
{\it Hamiltonian dynamics on the symplectic extended phase space for autonomous and non-autonomous 
systems} \cite{struck1}, the work done by Baldiotti and al. on {\it the quantization of the damped 
harmonic oscillator} \cite{baldiotti}, the work by Menouar and al. on {\it the quantization of the 
time-dependent singular potential systems: non-central potential in three dimension} \cite{menouar}
and the recent paper by Garcia-Chung and al. on the {\it Dirac's method for time-dependent Hamiltonian 
systems in the extended phase space} \cite{angel}.
In the extended phase space that is to consider the time $t$ as a dynamical variable with a corresponding 
conjugate momentum, the Lagrangian of the system is singular characterized then by the presence of constraints. 
We identify the constraints and apply the Dirac method of quantization.
This procedure has been presented in the paper by A. Garcia - Chung and al. \cite{angel}. 
An advantage of extending the phase space is that the symplectic group 
of the system is also enlarged giving place to study the canonical transformation in the extended phase space
such that the final dynamical description of the reduced phase space is no longer time-dependent. An invariant 
of the system can be obtained by applying in the extended phase space a finite canonical transformation to the 
initial Hamiltonian of the system.

The quantum Hamiltonian given in \cite{latevi} is straighforward through canonical quantization as the 
Lagrangian (\ref{lagrr}) is regular. Let's briefly recall here 
the procedure of canonical quantization of the system described in equations (\ref{lagrr},\ref{hamm},\ref{pbracket}). 
The Hessian matrix {\bf M} of the Lagrangian function is given by 

\begin{equation}
 {\bf M} = \left[
 \begin{array}{cc}
  \frac{\partial ^2 L}{\partial\dot{x}_1^2} & \frac{\partial}{\partial\dot{x}_1}[\frac{\partial L}{\partial\dot{x}_2}]\\
  \frac{\partial }{\partial\dot{x}_2}[\frac{\partial L}{\partial\dot{x}_1}] & \frac{\partial ^2 L}{\partial\dot{x}_2^2}
 \end{array} \right] =
 \left[ \begin{array}{cc}
                  f^{-1}m & 0 \\
                  0 & f^{-1}(m)
                 \end{array}\right]\;,
\end{equation}

and the determinant 

\begin{equation}\label{hess}
 \textrm{det}\; {\bf M} = \textrm{det}\;\left|\left| 
 \begin{array}{cc}
                  f^{-1}m & 0 \\
                  0 & f^{-1}(m)
                 \end{array}
  \right|\right| = m^2 f^{-2}(t)\neq 0\;.
\end{equation}

The Lagrangian in equation (\ref{lagrr}) is called regular or standard since its Hessian matrix satisfies the equation 
(\ref{hess}).  The system described by the Lagrangian in equation (\ref{lagrr}) does not involves constraints and we
assume that the phase space is flat and admits the procedure of
canonical quantization which consists in demanding that to the classical canonical pairs 
$(x_1,p_1),\; (x_2,p_2)$ that satisfy the Poisson brackets in equation (\ref{pbracket}) we associate the 
operators $\hat x_1,\; \hat x_2,\; \hat p_1,\; \hat p_2$ acting both on the Hilbert space of the states 
$\mathcal{H}$ and obey the canonical commutation relations 

\begin{equation}
[\hat x_1, \hat p_1] = i\hbar;\quad [\hat x_2, \hat p_2] = i\hbar;\quad
[\hat x_1, \hat p_2] = 0; \quad [\hat x_2,\hat p_1] =0\;, 
\end{equation}

where the commutator of two operators is given by $[\hat f, \hat g] = \hat f\hat g -\hat g\hat f$.
We assign to the classical Hamiltonian $H (x_1,x_2,p_1,p_2,t)$ in equation (\ref{hamm}) which is a 
function of the dynamical variables $ x_1,x_2,p_1,p_2 $ an operator $\hat H (\hat x_1, \hat x_2, \hat p_1, \hat p_2, t)$ 
which is obtained by replacing the dynamical variables with the corresponding operators.
Other classical dynamical quantities in Quantum Mechanics are similarly associated with quantum operators 
that act on the Hilbert space of states. 

The aim of this paper is to illuminate the Dirac's method of quantization of the system in the extended phase space.
The paper could interest some readers in the community of 
mathematical physics as it forms some integrity with all needed elements.
The organization of the paper is the following: in section \ref{sec2}, we apply the Dirac's 
method for constrained system to the model in the extended phase space. Concluding remarks are given in section \ref{sec3}.

\section{The model in the extended phase space}\label{sec2}

In this section we would like to perform the Dirac method's of quantization to 
constrained systems. The reader interested in knowing more about the method may check in 
the literature, for instance one may read 
in references \cite{dirac,faddeev,gitmann, girotti, jan, marc, klauder, glaure}. Here, a constrained system 
is one in which there exists a relationship between the system's degrees of freedom that 
holds for all times. 

We consider the time integral of the Lagrangian in equation (\ref{lagrr}) as the action

\begin{equation}\label{act}
 S\left[x_1(t),x_2(t),\frac{dx_1(t)}{dt},\frac{dx_2(t)}{dt} \right] = \int_{t_1}^{t_2}L(x_1,x_2,\dot x_1,\dot x_2)dt,
\end{equation}

In order to extend the phase space we consider the time parameter $t$ as an additional degree of freedom for the system $S$ described 
in equation (\ref{act}). We consider the arbitrary time scaling transformation $t = t_\tau (\tau)$, where the parameter $\tau$ 
plays the role of the new time parameter. The function $t_\tau (\tau)$ is chosen such that it gives a smooth 
one-to-one correspondence of the domain $\tau$ and $t$. This transformation changes also the dependency of the coordinates 
and requires the following redefinitions 

\begin{equation}
 x_1(t_\tau(\tau)) = x_{1,\tau}(\tau); \quad x_2(t_\tau(\tau)) = x_{2,\tau}(\tau),
\end{equation}

consequently we have a new functional expression for the action $S$ denoted by 

\begin{eqnarray}\nonumber
 S_\tau \left[x_{1,\tau},x_{2,\tau},t_\tau, \frac{dx_{1,\tau}}{d\tau},\frac{dx_{2,\tau}}{d\tau}, \frac{dt_\tau}{d\tau} \right]  &=& \int_{\tau_1}^{\tau_2} 
 \frac{m}{2}f^{-1}(t_\tau)\left(
 \left( \frac{\dot{x}_{1,\tau}}{\dot{t}_\tau}\right)^2 + \left(\frac{\dot{x}_{2,\tau}}{\dot{t}_\tau}\right)^2 \right.\\\label{acte}
 &-&\left. \frac{ m\omega^2(t_\tau)}{2}\left( x_{1,\tau}^2 + x_{2,\tau}^2\right) \right)\dot{t}_\tau d\tau,
\end{eqnarray}

where the notations $\dot{x}_{i,\tau}, \; i = 1,2$ and $\dot{t}_\tau$ are respectively 
$\displaystyle{\dot{x}_{i,\tau} = \frac{d}{d\tau} x_{i,\tau},\; i =1,2}$ and $\displaystyle{\dot{t}_\tau = \frac{d}{d\tau}{t_\tau}}$.
The generalized configuration variables on the extended phase space are given by $x_{1,\tau},\; x_{2,\tau},\; t_\tau$ and their velocities 
are respectively given by $\dot{x}_{1,\tau}, \dot{x}_{2,\tau}, \dot{t}_\tau$. We consider the boundary conditions 
$x_{1,\tau}(t_1) = x_1,\; x_{2,\tau}(t_2) = x_2,\; t_\tau(\tau_1) = t_1,\; t_\tau(\tau_2) = t_2 $.

The integrand of the equation (\ref{acte}) thus defines the extended Lagrangian $L_\tau$

\begin{equation}
 L_\tau (x_{1,\tau},x_{2,\tau},t_\tau,\dot{x}_{1,\tau},\dot{x}_{2,\tau},\dot{t}_\tau) = 
 \frac{f^{-1}(t_\tau) m}{2\dot{t}_\tau}\left(\dot{x}_{1,\tau}^2 + \dot{x}_{2,\tau}^2 \right) -\frac{m\omega^2(t_\tau)\dot{t}_\tau}{2}f^{-1}(t_\tau)
 \left(x_{1,\tau}^2 + x_{2,\tau}^2\right).
\end{equation}

Let's determine first the Hessian matrix ${\bf M}_\tau$ of the Lagrangian function $L_\tau$

\begin{equation}
 {\bf M}_\tau = \left[ 
 \begin{array}{ccc}
 \frac{\partial^2 L}{\partial \dot{x}_{1,\tau}^2} & 
 \frac{\partial^2 L}{\partial \dot{x}_{1,\tau}\partial \dot{x}_{2,\tau}} &
 \frac{\partial^2 L}{\partial \dot{x}_{1,\tau}\partial \dot{t}_\tau} \\
 {} & {} & {} \\
 \frac{\partial^2 L}{\partial \dot{x}_{2,\tau}\partial \dot{x}_{1,\tau}} &
 \frac{\partial^2 L}{\partial \dot{x}_{2,\tau}^2} & 
 \frac{\partial^2 L}{\partial \dot{x}_{2,\tau}\partial \dot{t}_\tau} \\
 {} & {} & {}  \\
 \frac{\partial^2 L}{\partial \dot{t}_\tau\partial \dot{x}_{1,\tau}} &
 \frac{\partial^2 L}{\partial \dot{t}_\tau\partial \dot{x}_{2,\tau}} &
 \frac{\partial^2 L}{\partial \dot{t}_\tau^2}
\end{array}
\right]\;,
\end{equation}

that is equivalent to

\begin{equation}
 {\bf M}_\tau = \left[ 
 \begin{array}{ccc}
  \frac{f^{-1}(t_\tau)m}{\dot{t}_\tau} & 0 & -\frac{f^{-1}(t_\tau)m \dot{x}_{1,\tau}}{\dot{t}_\tau^2}\\
  {} & {} & {} \\
  0 & \frac{f^{-1}(t_\tau)m}{\dot{t}_\tau} & -\frac{f^{-1}(t_\tau)m \dot{x}_{2,\tau}}{\dot{t}_\tau^2}\\
  {} & {} & {} \\
  -\frac{f^{-1}(t_\tau)m \dot{x}_{1,\tau}}{\dot{t}_\tau^2} & -\frac{f^{-1}(t_\tau)m \dot{x}_{2,\tau}}{\dot{t}_\tau^2} &
  \frac{f^{-1}(t_\tau)m (\dot{x}_{1,\tau}^2 +\dot{x}_{2,\tau}^2) }{\dot{t}_\tau^3}
 \end{array}
 \right]\;.
\end{equation}

It is easy to show that the determinant (Hessian) of the matrix ${\bf M}_\tau$ is zero, that means that 
the Lagrangian $L_\tau$ is singular and a singular Lagrangian theory necessarily involves constraints.
Let's determine now the corresponding conjugate momenta of the configuration variables $x_{1,\tau},\; x_{2,\tau},\; t_\tau$.
They are respectively given by

\begin{equation}
 p_{1,\tau} = \frac{\partial L_\tau}{\partial \dot{x}_{1,\tau}} = \frac{f^{-1}(t_\tau)m\dot{x}_{1,\tau}}{\dot{t}_\tau};
 \end{equation}
 
 \begin{equation}
 p_{2,\tau} = \frac{\partial L_\tau}{\partial \dot{x}_{2,\tau}} = \frac{f^{-1}(t_\tau)m\dot{x}_{2,\tau}}{\dot{t}_\tau};
 \end{equation}
 
 \begin{equation}
 p_\tau = \frac{\partial L_\tau}{\partial \dot{t}_\tau } = -\frac{f(t_\tau)}{2m}(p_{1,\tau}^2 + p_{2,\tau}^2)
 - \frac{m\omega^2(t_\tau)f^{-1}(t_\tau)}{2}
 (x_{1,\tau}^2 + x_{2,\tau}^2).
\end{equation}

 The momentum $p_\tau $ is expressed in term of the fundamental variables and a constraint arise as
 
\begin{equation}
 \phi  = p_\tau + \frac{f(t_\tau)}{2m}(p_{1,\tau}^2 + p_{2,\tau}^2)
 + \frac{m\omega^2(t_\tau)f^{-1}(t_\tau)}{2}
 (x_{1,\tau}^2 + x_{2,\tau}^2)\; \sim 0.
\end{equation}

We derive the extended Hamiltonian $H_\tau$ as the Legendre transform of the extended Lagrangian $L_\tau$

\begin{equation}
 H_\tau (x_{1,\tau},x_{2,\tau},t_\tau, p_{1,\tau},p_{2,\tau},p_\tau) = p_{1,\tau}\dot{x}_{1,\tau} 
 + p_{2,\tau}\dot{x}_{2,\tau} + p_\tau\dot{t}_\tau - L\;,
\end{equation}

that is explicitly

\begin{equation}
 H_\tau = \left( \frac{f(t_\tau)}{2m}(p_{1,\tau}^2 + p_{2,\tau}^2) + p_\tau
 + \frac{m\omega^2(t_\tau)f^{-1}(t_\tau)}{2}
 (x_{1,\tau}^2 + x_{2,\tau}^2) \right) \dot{t}_\tau.
\end{equation}

A first remark is that 

\begin{equation}
 H_\tau = \dot{t}_\tau \phi\; \sim 0\;.
\end{equation}

The use of $\sim $ sign instead of the $ = $ sign is due to Dirac \cite{dirac} 
and has a special meaning : two quantities related by a $\sim $ sign are only equal after all 
constraints have been imposed. Two such quantities are weakly equal to one another. It is 
important to note that the Poisson brackets in any expression must be worked out before any 
constraint are set to zero.

We have now an extended phase space determined by  $x_{1,\tau},\; x_{2,\tau},\; p_{1,\tau},\; p_{2,\tau},\; p_\tau$. 
The simplectic structure is determined by the non-vanishing Poisson Brackets 

\begin{eqnarray}\label{pbrac}
 \{x_{1,\tau}, p_{1,\tau}\}_{PB} = 1;\quad  \{x_{2,\tau}, p_{2,\tau}\}_{PB} = 1;\quad 
 \{t_\tau, p_\tau\}_{PB} = 1\;,
\end{eqnarray}

and the Poisson Brackets for two arbitrary smooth functions $f$ and $g$ in this extended phase space 
takes the following form 

\begin{equation}\label{pbrack}
 \{f,g\}_{PB} = \frac{\partial f}{\partial x_{1,\tau}}\frac{\partial g}{\partial p_{1,\tau}}
 + \frac{\partial f}{\partial x_{2,\tau}}\frac{\partial g}{\partial p_{2,\tau}} 
 + \frac{\partial f}{\partial t_\tau}\frac{\partial g}{\partial p_\tau}
 - \frac{\partial f}{\partial p_{1,\tau}}\frac{\partial g}{\partial x_{1,\tau}}
 -\frac{\partial f}{\partial p_{2,\tau}}\frac{\partial g}{\partial x_{2,\tau}}
 -\frac{\partial f}{\partial p_\tau}\frac{\partial g}{\partial t_\tau}
\end{equation}

The constraint $\phi$ is a primary constraint and the only one as there are no secondary constraints
generated. We have in presence a first-class constraint. Let's recall that a dynamical variable 
$R$ is said to be first-class if it has weakly vanishing Poisson brackets with all constraints.
The Hamiltonian $H_\tau$ is a first-class Hamiltonian. We set the total Hamiltonian to be 

\begin{equation}
 {H_{\tau}}_T = \lambda \phi\;,
\end{equation}

where $\lambda$ is a Lagrange multiplier, note that $\lambda $ depends only on time.
The Hamiltonian equations of motion derived with this Poisson bracket and the Hamiltonian ${H_{\tau}}_T$
are given by 

\begin{equation}
 \dot{x}_{1,\tau} =  \{x_{1,\tau}, {H_{\tau}}_T\}_{PB} = \lambda\frac{f(t_\tau)}{m}p_{1,\tau}; 
 \end{equation}
 
 \begin{equation}
 \dot{x}_{2,\tau} =  \{x_{2,\tau}, {H_{\tau}}_T\}_{PB}  = \lambda \frac{f(t_\tau)}{m} p_{2,\tau};
 \end{equation}
 
 \begin{equation}
 \dot{p}_{1,\tau} =  \{p_{1,\tau}, {H_{\tau}}_T\}_{PB} =  - \lambda m\omega^2(t_\tau)f^{-1}(t_\tau)x_{1,\tau};
 \end{equation}
 
 \begin{equation}
 \dot{p}_{2,\tau} = \{p_{2,\tau}, {H_{\tau}}_T\}_{PB} = - \lambda m\omega^2(t_\tau)f^{-1}(t_\tau)x_{2,\tau};
 \end{equation}
 
 \begin{equation}\label{elambda}
 \dot{t}_\tau = \{t_\tau, {H_{\tau}}_T\}_{PB} = \lambda;
 \end{equation}
 
 \begin{eqnarray}\nonumber
 \dot{p}_\tau &=& \{p_\tau, {H_{\tau}}_T\}_{PB};\\
 &=&  \lambda \left(
 \frac{\dot{f}(t_\tau)}{2m}(p_{1,\tau}^2 + p_{2,\tau}^2)
 + \frac{mf^{-1}(t_\tau)\omega(t_\tau)}{2}[2\dot\omega(t_\tau) -\omega(t_\tau)\dot{f}(t_\tau)f^{-1}(t_\tau)]
 (x_{1,\tau}^2 + x_{2,\tau}^2) \right).
\end{eqnarray}

The total Hamiltonian is proportionnal to the constraint $\phi$ and the coefficient of 
proportionality is a Lagrange multiplier denoted by $\lambda$. The Lagrange multiplier is independent 
of the phase space points. This kind of Lagrange multiplier is referred to as non-canonical gauge.
This particular case of constrained system in which the total Hamiltonian is null when the constraint 
is strongly set to zero are usually called reparametrization invariant system \cite{gitman}.
The fact that we have only first class constraint implies that all phase space functions will evolve 
by gauge transformations and the system at a given time will gauge equivalent to the system at any other 
time. To quantize such a theory we need to choose between the Dirac and the canonical quantization procedures.
If we choose the canonical quantization we face the fact that we have no Schr\"odinger equations because 
the total Hamiltonian must necessarily annihilate physical states. The solution is to impose a supplementary 
constraint $\eta$ that depends on the time variable. The process in which a value for the Lagrange multiplier $\lambda$ 
 is fixed is usually called {\it fixing the gauge} . For instance, the most 
 common gauge fixing is the case in which $\lambda = \frac{t_2 -t_1}{\tau_2 -\tau_1}$.
 This gauge solves the equation in (\ref{elambda}) $\dot{t}_\tau = \lambda $, that means 
 
 \begin{equation}
  t_\tau = \frac{(t_2 -t_1)(\tau -\tau_1)}{\tau_2 -\tau_1} +t_1\;,
 \end{equation}
 
with $t_\tau(\tau_i)  = t_i,\; i= 1,2 $ holds\;.  

The gauge fixing condition leads to an additional constraint surface 

\begin{equation}
 \eta = t_\tau - \frac{(t_2 -t_1)(\tau -\tau_1)}{\tau_2 -\tau_1} +t_1 \sim 0\;,
\end{equation}

and

\begin{equation}
\{\phi,\eta \} \nsim 0\;.
\end{equation}

The constraints $\phi$ and $\eta$ are second-class constraints. Let's recall that a 
dynamical variable $R$ is said be second-class if it has weakly non vanishing Poisson 
brackets with all the constraints. Let's define now the Dirac brackets. 
The matrix of the constraints is given by 

\begin{equation}
\Delta =
\left( 
 \begin{array}{cc}
  \{\phi,\phi\} & \{\phi,\eta\}\\
  \{\eta,\phi\} & \{\eta,\eta\}
 \end{array} \right) 
 = 
 \left( 
 \begin{array}{cc}
  0 & -1\\
  1 & 0
 \end{array} \right)\;,  
\end{equation}

the matrix $\Delta $ is obviously invertible  and its inverse is given by

\begin{equation}
 C = \left( \begin{array}{cc}
 0 & 1\\
 -1 & 0
\end{array}
\right)\;.
\end{equation}

The Dirac brackets of two extended phase space quantities $f$ and $g$ is given by 

\begin{equation}\label{dbrack}
 \{ f, g\}_{DB} = \{f,g\}_{PB} - \left[  
 \{f,\phi\}_{PB}\{\eta,g\}_{PB} - \{f,\eta\}_{PB}\{\phi,g\}_{PB}
  \right]\;.
\end{equation}

The Poisson bracket (\ref{pbrack}) is then replaced by the Dirac bracket (\ref{dbrack}).
With respect to that, let's calculate the Dirac brackets of the fundamental variables in the extended phase space.
The non vanishing Dirac brackets are the following

\begin{equation}\label{dbrac1}
 \{x_{1,\tau},p_{1,\tau}\}_{DB} =  1;
 \end{equation}
 
 \begin{equation}\label{dbrac2}
 \{x_{2,\tau},p_{2,\tau}\}_{DB} = 1;
 \end{equation}
 
 \begin{equation}\label{dbrac3}
 \{x_{1,\tau}, p_\tau \}_{DB}  =  -\frac{f(t_\tau)}{m}p_{1,\tau};
 \end{equation}
 
 \begin{equation}\label{dbrac4}
 \{x_{2,\tau}, p_\tau \}_{DB} = -\frac{f(t_\tau)}{m}p_{2,\tau};
 \end{equation}
 
 \begin{equation}\label{dbrac5}
 \{p_{1,\tau} , p_\tau \}_{DB} =  m\omega^2(t_\tau)f^{-1}(t_\tau)x_{1,\tau};
 \end{equation}
 
 \begin{equation}\label{dbrac6}
 \{p_{2,\tau} , p_\tau \}_{DB} =m\omega^2(t_\tau)f^{-1}(t_\tau)x_{2,\tau} .
\end{equation}

Comparing the Poisson brackets in equation (\ref{pbrac}) and the 
Dirac brackets in equations (\ref{dbrac1} - \ref{dbrac6}), we can note 
the differences that are essentially $\{t_\tau,\;p_\tau\}_{PB}  =1 $ while 
$\{t_\tau,\;p_\tau\}_{DB}  = 0$ and $\{p_{1,\tau} ,\; p_\tau \}_{PB} = \{p_{2,\tau} ,\; p_\tau \}_{PB}  = 0$ 
while $\{p_{1,\tau} ,\; p_\tau \}_{DB} =  m\omega^2(t_\tau)f^{-1}(t_\tau)x_{1,\tau}$ and \-
 $ \{p_{2,\tau} ,\; p_\tau \}_{DB} =m\omega^2(t_\tau)f^{-1}(t_\tau)x_{2,\tau}$ .
When the constraints are fulfilled that means $\phi = 0$  and $\eta =0$, we have 
the coordinates $x_{1,\tau},\; x_{2,\tau},\;p_{1,\tau},\;p_{2,\tau}$ selected as the  physical 
degree of freedom using $\tau$ as the time parameter or instead we can use $x_1,\;x_2,\;p_1,\;p_2$
with $t$ as time parameter in accordance with the initial description. In that situation, 
$t_\tau = \frac{(t_2- t_1)(\tau -\tau_1)}{\tau_2 -\tau_1} + t_1$ and 
$p_\tau = - H(x_{1,\tau},x_{2,\tau},p_{1,\tau},p_{2,\tau},\tau )$, where $H$ is the Hamiltonian 
in (\ref{hamm}). The dynamic of the system 
is then generated by the Hamiltonian $H$ and the non-vanishing 
Dirac brackets $ \{x_{1,\tau},\;p_{1,\tau}\}_{DB} =  1;\; \{x_{2,\tau},\;p_{2,\tau}\}_{DB} =  1;$

The canonical quantization procedure as described in section (\ref{sec1}) for an unconstrained system is to promote the phase 
space variable $x_1,\;x_2,\;p_1,\;p_2$ to operators $\hat{x}_1,\;\hat{x}_2,\;\hat{p}_1,\;\hat{p}_2$ that 
act on elements of a Hilbert space, which we denote $\vert \psi\rangle $. The commutator between 
phase space variables

\begin{equation}
 [\hat f, \hat g] = i\hbar \{f,g\}_{DB}\;, 
\end{equation}

and the quantum level Hamiltonian is given by 

\begin{eqnarray}
 \hat H = \frac{f(t)}{2m}(\hat{p}_1^2 +\hat{p}_2^2) + f^{-1}(t)\frac{m\omega^2(t)}{2}(\hat{x}_1^2 + \hat{x}_2^2),
\end{eqnarray}

with 

\begin{equation}
 [\hat{x}_1,\hat{p}_1] = i\hbar;\quad 
 [\hat{x}_2,\hat{p}_2] = i\hbar;\quad
 [\hat{x}_1,\hat{p}_2] = 0;\quad
 [\hat{x}_2,\hat{p}_1] = 0 .
\end{equation}

\section{Concluding remarks and perspectives}\label{sec3}

As we have already said in the introduction our aim in this paper is to 
illuminate the Dirac's quantization procedure for the model in the extended phase space. 
We focused then on the problem of Dirac's canonical quantization of two-dimensional, time-dependent 
harmonic oscillator. As result we showed that after performing all necessary steps the quantum 
system is simply modelled by the original quantum Hamiltonian.

The system can be studied as in \cite{latevi} by means of the Levis-Riesenfeld procedure of finding 
invariants hermitian operators. The invariant operator in \cite{latevi} is given by 

\begin{equation}
 \hat I(t) = \frac{1}{2}\left[
 (mf^{-1}\dot\rho\hat x_1 -\rho\hat p_1)^2 + \frac{\nu^2}{\rho^2}\hat x_1^2 + 
 (mf^{-1}\dot\rho\hat x_2 -\rho\hat p_2)^2 + \frac{\nu^2}{\rho^2}\hat x_2^2 \right],
\end{equation}

where the function $\rho$ is the solution of the so-called Ermakov-Pinney equation \cite{epiney}

\begin{equation}
 \ddot\rho +\eta\dot\rho +\omega^2\rho = \frac{\nu^2 f^2}{m^2\rho^3}\;.
\end{equation}

An alternative way of finding invariants of the system described in equations (\ref{lagrr},\ref{hamm}) is to study the canonical transformation 
in the extended phase space such that the final 
dynamical description of the reduced phase space is no longer time dependent. This method is discussed in 
\cite{angel}. The canonical transformation is a generalization of the Struckmeier transformation \cite{struck}.
For the present case we  consider a coordinate transformation of the form

\begin{equation}\label{ctransform}
\left[
 \begin{array}{cc}
  x_{1,\tau}\\ x_{2,\tau}\\ t_\tau \\p_{1,\tau}\\ p_{2,\tau}\\ p_\tau
 \end{array}\right] = 
 \left[ \begin{array}{cc}
 A_1(Q_1,T)\\ A_2(Q_2,T)\\B(T)\\C_1(Q_1,T)P_1 + D_1(Q_1,T)\\
 C_2(Q_2,T)P_2 + D_2(Q_2,T)\\
 F(Q_1,Q_2,T,P_1,P_2,P_T)
 \end{array}
 \right]\;.
\end{equation}

The canonical transformation matrix resulting from (\ref{ctransform}) is given by 

\begin{equation}
 \mathcal{M} = \left[
 \begin{array}{cccccc}
  \frac{\partial x_{1,\tau}}{\partial Q_1} & \frac{\partial x_{1,\tau}}{\partial Q_2} & \frac{\partial x_{1,\tau}}{\partial T} &
  \frac{\partial x_{1,\tau}}{\partial P_1} & \frac{\partial x_{1,\tau}}{\partial P_2} & \frac{\partial x_{1,\tau}}{\partial P_T}\\
  {}\\
  \frac{\partial x_{2,\tau}}{\partial Q_1} & \frac{\partial x_{2,\tau}}{\partial Q_2} & \frac{\partial x_{2,\tau}}{\partial T} &
  \frac{\partial x_{2,\tau}}{\partial P_1} & \frac{\partial x_{2,\tau}}{\partial P_2} & \frac{\partial x_{2,\tau}}{\partial P_T}\\
  {}\\
  \frac{\partial t_\tau}{\partial Q_1} & \frac{\partial t_\tau}{\partial Q_2} & \frac{\partial t_\tau}{\partial T} &
  \frac{\partial t_\tau}{\partial P_1} & \frac{\partial t_\tau}{\partial P_2} & \frac{\partial t_\tau}{\partial P_T}\\
  {}\\
  \frac{\partial p_{1,\tau}}{\partial Q_1} & \frac{\partial p_{1,\tau}}{\partial Q_2} & \frac{\partial p_{1,\tau}}{\partial T} &
  \frac{\partial p_{1,\tau}}{\partial P_1} & \frac{\partial p_{1,\tau}}{\partial P_2} & \frac{\partial p_{1,\tau}}{\partial P_T}\\
  {}\\
  \frac{\partial p_{2,\tau}}{\partial Q_1} & \frac{\partial p_{2,\tau}}{\partial Q_2} & \frac{\partial p_{2,\tau}}{\partial T} &
  \frac{\partial p_{2,\tau}}{\partial P_1} & \frac{\partial p_{2,\tau}}{\partial P_2} & \frac{\partial p_{2,\tau}}{\partial P_T}\\
  {}\\
  \frac{\partial F}{\partial Q_1} & \frac{\partial F}{\partial Q_2} & \frac{\partial F}{\partial T} &
  \frac{\partial F}{\partial P_1} & \frac{\partial F}{\partial P_2} & \frac{\partial F}{\partial P_T}
 \end{array}
 \right]\;,
 \end{equation}
 
 that is equivalent to 
 
 \begin{equation}
 \mathcal{M} = 
 \left[\begin{array}{cccccc}
  A'_1 & 0 & \dot{A}_1 & 0 & 0 & 0\\
  {}\\
  0 & A'_2 & \dot{A}_2 & 0 & 0 & 0\\
  {}\\
  0 & 0 & \dot{B} & 0 & 0 & 0 \\
  {}\\
  C_1'P_1 +D_1' & 0 & \dot{C}_1P_1 +\dot{D}_1 & C_1 & 0 & 0\\
  {}\\
  0 & C'_2P_2 + D'_2 & \dot{C}_2 P_2 + \dot{D}_2 & 0 & C_2 & 0\\
  {}\\
  \frac{\partial F}{\partial Q_1} & \frac{\partial F}{\partial Q_2} & \frac{\partial F}{\partial T} & 
  \frac{\partial F}{\partial P_1} & \frac{\partial F}{\partial P_2} & \frac{\partial F}{\partial P_T}
 \end{array}\right]\;,
\end{equation}

where $A'_i = \frac{\partial A_i}{\partial Q_i};\quad C'_i = \frac{\partial C_i}{\partial Q_i};\quad D'_i = \frac{\partial D_i}{\partial Q_i}, \quad  i=1,2$
and the {\it dot} notation is used for the derivative with respect to T.
We would like to solve $\mathcal{M}^T J \mathcal{M} = J$ where $J$ is the matrix 

\begin{equation}
 J = \left[
 \begin{array}{cccccc}
  0 & 0 & 0 & 1 & 0 & 0\\
  0 & 0 & 0 & 0 & 1 & 0 \\
  0 & 0 & 0 & 0 & 0 & 1\\
  -1 & 0 & 0 & 0 & 0 & 0\\
  0 & -1 & 0 & 0 & 0 & 0 \\
  0 & 0 & -1 & 0 & 0 & 0
 \end{array}
\right]\;.
\end{equation}

Solving $ \mathcal{M}^T J \mathcal{M} = J $ leads to a system of differential equations

\begin{equation}
 \dot{A}_1(C_1'P_1 + D_1') + \dot{B}\frac{\partial F}{\partial Q_1} = (\dot{C}_1 P_1 +\dot{D}_1)A'_1;
\end{equation}

\begin{equation}
 \dot{A}_2(C_2'P_2 + D_2') + \dot{B}\frac{\partial F}{\partial Q_2} = (\dot{C}_2 P_2 +\dot{D}_2)A'_2;
\end{equation}
 
\begin{equation}
 C_1\dot{A}_1 +\dot{B}\frac{\partial F}{\partial P_1} = 0;
\end{equation}
 
\begin{equation} 
 C_2\dot{A}_2 +\dot{B}\frac{\partial F}{\partial P_2} = 0;
\end{equation}
 
\begin{equation} 
 \dot{B}\frac{\partial F}{\partial P_T}  = 1;
\end{equation}
 
\begin{equation}
 C_1A_1' = 1;
 \end{equation}
 
 \begin{equation}
 C_2A_2' = 1\;
 \end{equation}
 
whose general solution is given by 

\begin{equation}
 C_1 = \frac{1}{A'_1}; \quad C_2 = \frac{1}{A_2'};\quad 
 t_\tau = B(T);
 \end{equation}
 
 \begin{equation}
 F =  \frac{P_T}{\dot{B}} - \frac{\dot{A}_1}{A_1'\dot{B}}P_1 
 -\frac{\dot{A}_2}{A_2'\dot{B}}P_2 + \frac{1}{\dot{B}}
 \left[\int\left(\dot{D}_1A_1' -\dot{A}_1D_1'\right)dQ_1 
 + \int\left(\dot{D}_1A_2' -\dot{A}_2D_2'\right)dQ_2 \right]\;,
\end{equation}

where the functions $A_i(Q_i,T),\; B(T),\;D_i(Q_i,T),\; i = 1,2\;$ are arbitrary. We can now write 
$x_{1,\tau},\;x_{2,\tau},\;t_\tau,\; p_{1,\tau},\;p_{2,\tau},\;p_\tau $ in terms of the new coordinates as 

\begin{equation}
\left[
 \begin{array}{cc}
  x_{1,\tau}\\ x_{2,\tau}\\ t_\tau \\p_{1,\tau}\\ p_{2,\tau}\\ p_\tau
 \end{array}\right] = 
\left[
\begin{array}{c}
 A_1(Q,T)\\
 A_2(Q,T)\\
 B(T)\\
 \frac{1}{A_1'}P_1 + D_1\\
 \frac{1}{A_2'}P_2 + D_2\\
 \frac{P_T}{\dot{B}} - \frac{\dot{A}_1}{A_1'\dot{B}}P_1 
 -\frac{\dot{A}_2}{A_2'\dot{B}}P_2 + \frac{1}{\dot{B}}
 \left[\int\left(\dot{D}_1A_1' -\dot{A}_1D_1'\right)dQ_1 
 + \int\left(\dot{D}_1A_2' -\dot{A}_2D_2'\right)dQ_2 \right]
\end{array}
\right]\;.
\end{equation}

The new variables are time independent since the time variable is $\tau$. A new Hamiltonian 
of the system can be derived in terms of these new variables that is also an invariant of the system
since autonomous.

\noindent {\bf Acknowledgment} : L. Gouba would like to thank the reviewers for their comments and suggestions.

\end{document}